\documentclass{appolb}
\usepackage{epsfig}

\begin{document}
\pagestyle{plain}
\title{ Unitarity corrections from the high energy QCD effective action 
\thanks{Presented at the "School on QCD, Low x Physics, Saturation and Diffraction", Copanello, Italy, July 2007}
}
\author{Martin Hentschinski 
\thanks{Work in collaboration with J. Bartels and L.N. Lipatov} 
\address{II. Institut f\"ur Theoretische Physik, Universit\"at Hamburg, Luruper Chaussee 149, 22761 Hamburg, Germany}
}
\maketitle
\begin{abstract}
  We investigate the derivation of reggeon transition vertices from
  Lipatov's gauge invariant effective action for high energy processes in QCD. In
  particular we address the question of longitudinal integrations in
  order to reduce the vertices into the required purely transverse
  form. We explicitly derive the BFKL-kernel and verify vanishing of
  the 2-to-3 reggeon transition vertex. First results on the
  derivation of the 2-to-4 reggeon transition vertex are discussed.
\end{abstract}
  
\section{Introduction}
In 1995 an effective action \cite{lipatov, antonov} for QCD scattering processes at high
center of mass energies has been proposed by L.N. Lipatov which
describes the interaction of reggeized gluons with quark-( $\psi$) and
gluon-($v_\mu$) fields local in rapidity. The action is formulated for the
Quasi-Multi-Regge-Kinematics (QMRK), where quark and gluon fields
build clusters  localized in rapidity, while these clusters
themselves are ordered strongly in rapidity w.r.t each other. Their
overall rapidity is bounded from above and below by the rapidities of the
scattering particles which are taken to propagate along opposite light-cone directions $n^+$ and $n^-$. Interaction between the clusters is mediated by
the reggeon fields $A_\pm$. The  Lagrangian of the effective action  is given by 
\begin{eqnarray}
  \label{eq:effact1}
  \mathcal{L}_{\mbox{eff}} 
&=&
  \mathcal{L}_{\mbox{QCD}} (v_\mu, \psi) + 
(A_-(v_\mu) -A_-)\partial^2 A_+ + (A_+(v_\mu) -A_+)\partial^2 A_-
 \nonumber \\
 \mbox{with} & A_\pm(v)&  =v_\pm D_\pm^{-1}\partial_\pm = 
v_\pm - gv_\pm\frac{1}{\partial_\pm}v_\pm + g^2v_\pm\frac{1}{\partial_\pm}v_\pm\frac{1}{\partial_\pm}v_\pm - \ldots
\end{eqnarray}
Light-cone components are defined by $k^\pm \equiv n^\pm\cdot k$.  To
obtain quantities of interest from the action, it is necessary to
perform integrations over light-cone degrees of freedom, which in
particular requires to find the right prescription for the poles in
$\partial_\pm$.  Formally these poles are harmless as the action is
given for central rapidities, but to obtain reliable results, a solid
understanding of these poles and the integration over light-cone
components in general is crucial. In order to do so, we first attempt to
rederive some well-known results from the action. In the following we
present some first results.

\section{The gluon trajectory}
\label{sec:traj}

We start  with the gluon trajectory. From the Feynman rules of the effective action \cite{antonov}  one obtains 
\begin{eqnarray}
  \label{eq:traj1}
  2 g^2 N_c \int \frac{dk^+ dk^- d^2 {\bf{k}}}{(2 \pi)^4} \frac{{\bf{q}}^2}{k^+}  \frac{{\bf{q}}^2}{k^-} \frac{-i}{k^+k^- - {\bf{k}}^2 + i \epsilon}   \frac{-i}{k^+k^- - ({\bf{k}} - {\bf{q}})^2 + i \epsilon}.
\end{eqnarray} 
Comparing this  with the corresponding one-loop QCD amplitude in the Regge limit, it appears that the poles in $k^+$ and $k^-$ should be interpreted as a principal value i.e. $1/k^- = 1/2 (1/(k^- + i\epsilon) + 1/(k^- - i\epsilon) )$. Performing the integral over $k^-$ by closing the contour at infinity one finds
\begin{eqnarray}
  \label{eq:traj_res}
 (-i 2 {\bf q}^2)  \int \frac{d k^+}{|k^+|}        \beta({\bf q}),  && \beta({\bf q})=   g^2  \frac{N_c}{2} \int \frac{d^2 {\bf k}}{(2 \pi)^3} \frac{- {\bf q}^2}{{\bf k}^2 ({\bf q } - {\bf k})^2 },
\end{eqnarray}
where the $k^+$ integral will produce the logarithm in $s$.

\section{The 2-to-2 transition kernel}
\label{sec:222ker}

\begin{figure}[htbp]
  \centering
  \includegraphics[width = 4cm]{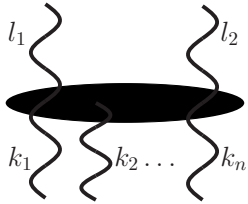}
  \caption{Transition of $2$ to $n$ reggeons.  For all transition kernels,  momenta and color labels  from above will be  denoted by $l$ and $a$ respectively, while those from below  by $k$ and $b$ respectively, with $ {\bf q} = \sum_i {\bf k_i}  = \sum_j {\bf l_j}$,  and ${\bf q}^2 = -t$ being the momentum transfer}
  \label{fig:222ker}
\end{figure}

Next we come to the 2-to-2 reggeon transition kernel (see fig.
\ref{fig:222ker}) which is the real part of the BFKL-kernel.  From
the effective action one obtains:
\begin{eqnarray}
  \label{eq:bfkl1}
  && i8g^2\int \frac{d l^+ d k^- }{2\pi} \bigg[ \frac{f^{a_1b_1c}f^{cb_2a_2}}{ -l^+ k^-  -({ \bf  k} - {\bf l})^2 + i\epsilon} ( -{{\bf q}}^2 - \frac{({ \bf  l} -{{ \bf q}})^2 {{\bf k}}^2}{l^+ k^-} - \frac{({\bf  k} -{\bf q})^2 {{\bf l}}^2}{l^+ k^-}) 
\nonumber \\
&& +  \frac{ f^{a_2b_1c}f^{cb_2a_1}}{l^+k^- -({\bf q} -{\bf k} -{\bf l})^2 + i\epsilon}     ( -{\bf q}^2 + \frac{{\bf l }^2 {\bf k}^2}{l^+ k^-} 
+
  \frac{({\bf k} -{\bf q})^2 ({ \bf l} -{\bf q})^2}{l^+k^-})\bigg].
\end{eqnarray}
Attempting to perform the integral over $k^-$ by closing the contour at infinity, one faces the problem that the term proportional to ${{\bf
    q}}^2$ is logarithmically divergent. However, if one symmetrizes in the color labels $b_1$ and $b_2$,
 the singularity cancels and the contour can be closed. Together with the remaining terms (where poles in light-cone momenta are interpreted as in section \ref{sec:traj}) one obtains the real (connected) part of the BFKL-Kernel. In the anti-symmetric case however, the singularity does not cancel. Instead (defining $\mu = -l^+ k^-$) we face  the following integral 
\begin{eqnarray}
  \label{eq:asympv}
 \int \frac{d l^+}{|l^+|} \int \frac{d \mu}{2 \pi} (\frac{1 }{ \mu - ({\bf k} - {\bf l})^2 + i \epsilon}- \frac{1 }{- \mu - ({\bf q } - {\bf k} - {\bf l})^2 + i \epsilon}).
\end{eqnarray}
As the effective theory is formulated for central rapidities, the
integration over $\mu$ can be understood as $\int d\mu := \lim_{M \to
  \infty} \int_{-M}^M d\mu$ which  leads to  vanishing of
(\ref{eq:asympv}). This might appear surprising from the first sight,
as usually the BFKL-Kernel with anti-symmetric color configuration is
associated with reggeization of the gluon. However,  in the
effective theory reggeization is  taken into account by
resuming contributions like (\ref{eq:traj_res}), and therefore vanishing of (\ref{eq:asympv}) is needed in order to avoid double
counting.

\section{The 2-to-3 transition}
\label{sec:223ker}

Due to signature conservation, the 2-to-3 transition has to vanish. For the connected part of the kernel one obtains  for a particular ordering of color and momenta indices 
\begin{eqnarray}
  \label{eq:threetrans}
 & &2^5 g^3 f^{a_1b_1c_1}f^{c_1b_2c_2}f^{c_2b_3a_2} \int \frac{dl^+}{l^+} \int \frac{d\mu_1 d\mu_3}{(2\pi)^3}
\big[( -{{\bf q}}^2 
 +
 \frac{({{\bf q}\! -\!{\bf l}})^2({\bf k_1}\! +\!{ \bf k_2})^2}{-\mu_3 } +
\nonumber \\
&&\frac{{{\bf l}}^2({\bf k_2}\! +\! {\bf k_3})^2}{\mu_1} 
\! -\! 
 \frac{ {{\bf l}}^2 {{\bf k_2}}^2 ({{\bf q}\! -\!{\bf l}})^2 }{\mu_1(-\! \mu_3)}
) 
\frac{1}{(\mu_1\! -\! ({\bf l}\! -\! {\bf k_1})^2\!+\!i\epsilon)(-\mu_3\! -\! ({\bf  l}\! -\! {\bf k_1}\! -\! {\bf k_2})^2\! +\! i\epsilon)}
\nonumber \\
&&+
\frac{{{\bf l}}^2 {{\bf k_3}}^2}{\mu_1(-\mu_3)} \frac{1}{\mu_1\! -\! ({\bf l}\! -\! {\bf k_1})^2\!+\!i\epsilon } 
 +
\frac{({{\bf q}\!-\!{\bf l}})^2 {{\bf k_1}}^2}{\mu_1(-\mu_3)} \frac{1}{-\mu_3 -\! ({\bf l}\! -\! {\bf k_1}\!-\!{\bf k_2})^2\!+\!i\epsilon }
\big],
\end{eqnarray}
where $\mu_i = -l^+k_i^-$, $i = 1,2,3 $ and $\mu_1 + \mu_2 + \mu_3
=0$. The above expression needs to be supplemented by expressions
which arise due to all permutations of the indices 1,2,3.
Applying the above argument we define (the formally divergent) integral
over $l^+$ by $\int dl^+ := \lim_{\Lambda \to \infty}
\int_{-\Lambda}^\Lambda dl^+$ which puts (\ref{eq:threetrans}) to
zero. A similar results holds for the disconnected pieces.

\section{The 2-to-4 transition vertex}
\label{sec:224trans}

The 2-to-4 transition is allowed by signature conservation. A building
block of the complete 2-to-4 kernel, namely the Reggeon-Particle-2
Reggeons vertex, was derived from the effective action and brought
into a purely transverse form in \cite{braun}.  The complete
unintegrated 2-to-4 vertex can be derived from (\ref{eq:effact1}), but
the expression is too lengthy to be reproduced here.  In contrast to
(\ref{eq:threetrans}), the $l^+$ integral is replaced by $\int d
l^+/|l^+|$ and therefore the integral does not vanish by the previous
argument, but leads to a logarithm in $s$.  The integrals over the
variables $\mu_i = -l^+ k_i^-$, $i=1,2,3,4$ can be performed using the
principal value pole prescription, but the obtained result does not
reproduce correctly all parts of the 2-to-4 vertex as derived in
\cite{bartelswue}. Therefore the pole prescription in terms of
principal values needs to be generalized in a way that it reproduces
all energy discontinuities of the 2-to-4 reggeon vertex. This is work
in progress.

\section{Conclusions}
\label{sec:concl}
Using the principal value pole prescription we rederived the gluon
trajectory and reproduced the 2-to-2 kernel in the symmetric color
configuration. The anti-symmetric color configuration was shown to
vanish if one takes into account that the effective action is
formulated for central rapidities.  By the same argument the 2-to-3
transition vanishes.  To integrate out the longitudinal degrees of
freedom of the 2-to-4 transition kernel, it seems to be necessary to
generalize the principal value pole prescription to reproduce
correctly all energy discontinuities.

\section*{Acknowledgements}
This work was supported by the Graduiertenkolleg "Zuk\"unftige
Entwicklungen in der Teilchenphysik"

\end{document}